\documentclass[12pt]{article}
\usepackage[margin=2.5cm]{geometry}
\usepackage[utf8]{inputenc}
\usepackage{amsmath}
\usepackage{amssymb}
\usepackage{amsthm}
\usepackage{enumitem}
\usepackage{hyperref}
\usepackage{graphicx}
\usepackage{colortbl}
\usepackage{multirow}
\usepackage{array}
\usepackage{siunitx}
\usepackage{xr}

\usepackage{lineno}

\usepackage{natbib}
\usepackage{graphicx}

\usepackage{url}

\usepackage{pdflscape}
\usepackage{afterpage}

\usepackage{array}
\usepackage{xr}

\title{PEtab-GUI: A graphical user interface to create, edit and inspect PEtab parameter estimation problems}
\author{
Paul Jonas Jost\,$^{\text{1,2}}$,
Frank T Bergmann\,$^{\text{3}}$,
Daniel Weindl\,$^{\text{1,2}}$,
Jan Hasenauer\,$^{\text{1,2,}\ast}$
}
\date{}

\begin{document}

\maketitle
{\small
$^{\text{1}}$ Bonn Center for Mathematical Life Sciences, University of Bonn, Bonn, Germany\\
$^{\text{2}}$ LIMES Life and Medical Sciences Institute, University of Bonn, Bonn, Germany\\
$^{\text{3}}$ BioQUANT, Heidelberg University, Heidelberg, Germany\\
$^{\text{*}}$ Lead contact: jan.hasenauer@uni-bonn.de\\
}

\abstract{
\textbf{Motivation:} Parameter estimation is a cornerstone of data-driven modeling in systems biology. Yet, constructing such problems in a reproducible and accessible manner remains challenging. The PEtab format has established itself as a powerful community standard to encode parameter estimation problems, promoting interoperability and reusability. However, its reliance on multiple interlinked files — often edited manually — can introduce inconsistencies, and new users often struggle to navigate them. Here, we present PEtab-GUI, an open-source Python application designed to streamline the creation, editing, and validation of PEtab problems through an intuitive graphical user interface. PEtab-GUI integrates all PEtab components, including SBML models and tabular files, into a single environment with live error-checking and customizable defaults. Interactive visualization and simulation capabilities enable users to inspect the relationship between the model and the data. PEtab-GUI lowers the barrier to entry for specifying standardized parameter estimation problems, making dynamic modeling more accessible, especially in educational and interdisciplinary settings.

\textbf{Availability and Implementation:} PEtab-GUI is implemented in Python, open-source under a 3-Clause BSD license. The code, designed to be modular and extensible, is hosted on \href{https://github.com/PEtab-dev/PEtab-GUI}{https://github.com/PEtab-dev/PEtab-GUI} and can be installed from \href{https://pypi.org/project/PEtab-GUI/}{PyPI}.

}

\section{Introduction}

Data-driven mathematical models are integral to understanding complex biological systems and processes. Yet, parameters of such models often remain unknown, necessitating parameter estimation \citep{VillaverdePat2021, ArmisteadHoe2024, BurbanoKem2024}. Established and calibrated data-driven mathematical models provide a key resource in further investigation of biological systems, from experimental design to model prediction \citep{HassLoo2019, MalikGlo2020}. 

Parameter estimation is nowadays supported by a growing spectrum of tools. However, workflows are often fragmented and difficult to reproduce or adapt \citep{TiwariKri2021}.
Indeed, until the introduction of the parameter estimation table (PEtab) format \citep{SchmiesterSch2021}, most tools came with their own input formats, making it harder to switch between them and to reproduce or reuse results. PEtab has successfully bridged the gap between multiple such toolboxes, vastly improving the reproducibility and reusability of parameter estimation problems in systems biology. It is supported by multiple toolboxes across various programming languages such as \citep{HoopsSah2006, RaueSte2015, FroehlichWei2021, SchaelteFro2023, PerssonFro2025}.

PEtab enables the specification of parameter estimation problems based on standardized tabular files and model specification standards (i.e. SBML \citep{HuckaFin2003}). 
Yet, despite its many advantages, PEtab’s reliance on a coordinated set of interrelated tabular and model files can itself be a source of error for inexperienced users. For example, renaming an observable identifier will have implications on three of five tabular files, warranting changes there. Additionally, allowed table attributes may not be known by heart, slowing down the editing process. Keeping track of all interconnections manually while constructing the PEtab files separately through conventional editors (e.g., Microsoft Excel, Numbers) makes the process arduous and error-prone. Programmatically generating PEtab files poses a high entry barrier, since it demands both an understanding of the file structure and the programming skills to translate problems into the PEtab format. Thus, there is a need to improve the accessibility of the PEtab format, facilitating entry to data-driven mathematical modeling and parameter estimation for a larger group of scientists. 

Here, we present PEtab-GUI, a Python-based graphical user interface to aid in creating, editing, and inspecting parameter estimation problems defined in the PEtab format. The application allows combined editing of all the PEtab tabular files as well as the SBML file. PEtab-GUI automates the tracking of the aforementioned interconnections. It allows checking the validity of the current state of the PEtab problem, instantaneous checks for edited cells, and inspection of the integrated data. PEtab-GUI is broadly applicable but particularly valuable for scientists new to dynamic modelling and parameter estimation using PEtab. Thus, our application makes PEtab more accessible and provides a further entry point to data-driven mathematical modelling.

\section{Features}\label{sec:Features} 

PEtab encodes parameter estimation problems using a collection of files. Following the original definition \citep{SchmiesterSch2021}, these files are:
the \textbf{model file} describes the biological system using the SBML format, which is widely supported and allows reusing existing models without modification. The \textbf{condition file} encodes the experimental settings, such as treatments or genetic backgrounds, under which datasets are collected. The \textbf{observable file} maps internal model variables to measurable outputs via observation functions. It also defines noise distributions (e.g., normal or Laplace) and their parameters, which can be estimated as part of the modeling process. The \textbf{measurement file} contains the actual experimental data and links each data point to its corresponding condition and observable. It supports optional pre-equilibration settings and allows for measurement-specific overrides of observation parameters. One defines the parameters to be estimated and their bounds in the \textbf{parameter file}. It can also include prior distributions for use in optimization or Bayesian inference. Optionally, one can specify how to display data and simulation results together, such as time-course or dose-response plots, in the \textbf{visualization file}. Combining all previously mentioned files, the \textbf{PEtab problem file} serves as a central configuration linking all individual PEtab components. It enables flexible combinations of files for reuse or model comparison, using YAML format.

To facilitate the creation, editing, and inspection of PEtab problems, we designed PEtab-GUI as a graphical user interface that aids both new and experienced modelers in systems biology. We developed PEtab-GUI according to user feedback and have additionally tested it with problems of various sizes from a benchmark collection \citep{HassLoo2019}. The PEtab-GUI features are organized roughly into the aforementioned three major categories, each addressing key bottlenecks in the creation, editing, and inspection of parameter estimation problems.

\subsection{Opening, Creating and Archiving}

The application supports opening both complete and incomplete problems and individual tables, which can also be dragged and dropped into the interface. It automatically detects the type of file users open and integrates it seamlessly into the current parameter estimation problem. It also resolves a common formatting mismatch between experimental data and the PEtab standard. While PEtab requires each measurement (a specific observable measured at a specific time under specific conditions) to be entered as a separate row, experimental data is often stored in matrix format, where rows correspond to timepoints or doses and columns to observables. PEtab-GUI allows the user to import such matrices directly, transforming them into the required PEtab format. In addition, it automatically creates a template for the necessary observables and conditions, ensuring the problem remains PEtab-compliant without requiring manual restructuring. The user can save every table and the SBML model individually, or the whole PEtab problem can be saved to a directory or as a COMBINE archive \citep{BergmannAda2014}.

\subsection{Interactive and Intuitive Editing}

One of the biggest obstacles in creating a PEtab problem from scratch is the need to create up to five different interdependent tables. PEtab-GUI resolves this issue by designing every data table as a dock widget, a freely resizable and movable window, whose visibility can be toggled, allowing the user to move single tables into separate screens. Each table supports intuitive operations such as adding or deleting rows and columns, and copying and pasting content. To reduce manual effort and potential errors, PEtab-GUI provides context-aware features. Fields with predefined valid entries, such as the ``estimate'' column in the parameter table, rely on combo-boxes to guide selection. Other columns, such as the parameterId in the parameter table, whose options are dynamically specified by the other PEtab components, offer drop-down menus upon editing. PEtab table columns commonly have duplicate values. To speed up the editing process, PEtab-GUI allows editing multiple cells in the same column at once, giving each the same value. Moreover, when a user references a new condition or observable in the measurement table, the software automatically generates a corresponding entry with customizable default values in the appropriate table. The software also includes live cell validation: as users modify entries, the software performs real-time plausibility checks using PEtab-compliant linting tools. This helps catch structural inconsistencies or incomplete data early in the modeling process. Additional conveniences include advanced table filtering and the use of model-derived defaults for values such as parameter bounds or noise specifications.

The SBML model can be viewed in a separate tab, either as the original XML code or auto-converted to the Antimony format \citep{SmithBer2009}, which uses a concise and human-readable syntax for specifying reaction networks. The user can view and edit both versions that are synchronized with each other.

\subsection{Visualization, Simulation and Validation}

Visualization, simulation, and validation are tightly integrated into the modeling workflow. PEtab-GUI visualizes both, measurement and simulation data, facilitating visual validation. The plots can be generated based on different experimental conditions, outputs, or individually through the visualization specification table. The visualization also supports bidirectional highlighting: when users select a data point or condition in the plot, the corresponding entry is highlighted in the table, and vice versa, allowing for seamless navigation between data and model. Since live updates of multiple plots can be computationally demanding, we tested a variety of problems from the PEtab-Benchmark collection. The application was able to open all problems, though it did take a considerable amount of time for tables with several thousand rows. PEtab problems with about 50 parameters and 300 measurements, which covers the majority of current PEtab problems, were opened and, without noticeable lag, edited and scanned for errors. However, bigger models got slower, especially due to the extensive time spent on plotting and live filtering. We implemented the possibility to turn off the plotting by simply making the plotting widget invisible. We therefore deem the application, with all its features active, adequately usable up to medium-sized models. Turning off certain features can still allow larger models to be edited without too much lag.

To assess model behavior in real time, users can trigger simulations directly within the interface. A single click simulates the current model with BasiCO \citep{Bergmann2023} using the latest parameter values, and the system automatically overlays results on the measurement data. This integration provides immediate visual feedback, making it easy to check for available literature values and general model structure.

\section{Software Implementation} \label{sec:SoftwareImplement}

The application is written in Python 3.11 and installable from PyPI. It is built upon PySide6 \citep{Fitzpatrick2021}, the official Qt binding for Python. This choice offers a robust foundation for cross-platform compatibility while providing the flexibility needed to design an intuitive and responsive user experience tailored to the needs of PEtab problem creation. The code is structured in a Model-View-Controller (MVC) architecture to separate data management, user interface and interaction logic. Each of the three components in turn is hierarchically structured for each of the application's components, such that each component on its own is still using the MVC architecture. This approach not only enhances maintainability and readability but also enables scalable development, as new features can be added with minimal interference to existing components.

\section{Discussion}

A central aim of PEtab-GUI is to lower the barrier to entry for systems biology modeling by facilitating the construction of standardized parameter estimation problems. Modelling pipelines often require fluency in multiple file formats, coding languages, and simulation tools. By contrast, this application integrates creation, editing, and inspection into an intuitive graphical user interface, enabling users to build parameter estimation problems without requiring extensive programming knowledge or frequent switching between applications. This makes the tool particularly well-suited as an entry point to data-driven mathematical modeling and for educational settings.

Ongoing maintenance plays a key role in determining a tool's long-term impact and thus its prolonged survival in the scientific community. Using a modular and hierarchical structure not only reduces the maintenance effort but also lays a foundation for future feature expansion. This also includes the necessary extension to future PEtab versions, keeping the support up to date.

A possible extension involves integrating parameter estimation execution, a predominantly programmatic task. This mainly includes the integration of optimizer and simulator choices, as well as evaluation options. A crucial step in this endeavor involves enabling optimization, a cost-intensive task, to be run on clusters.

While the GUI-centered design provides clarity and structure, it may also impose limits on flexibility when compared to fully script-based modeling. For highly specialized workflows or non-standard use cases, advanced users may still prefer direct code-level access to PEtab or SBML. This might also be the case when handling large-scale models. The reactivity of the PEtab-GUI is one of its main selling points, but it has the drawback of being cost-intensive. 

A key strength of the tool lies in its built-in, real-time validation of user inputs. By integrating PEtab linting directly into the table editors, the application shifts error detection to the earliest stages of model development. This proactive approach minimizes silent inconsistencies and encourages best practices throughout the modeling process. It also ensures full compatibility with downstream toolchains, leveraging the PEtab standard to enable smooth and reliable workflows.

\section{Competing interests}
No competing interest is declared.

\section{Author contributions statement}

P.J.J., F.T.B. and D.W. developed the software.  P.J.J., F.T.B., D.W. and J.H. wrote and reviewed the initial draft of the manuscript. J.H. supervised the project and provided funding.

\section{Acknowledgments}

This work was supported by the Deutsche Forschungsgemeinschaft (DFG, German Research Foundation) under Germany’s Excellence Strategy (EXC 2047—390685813, EXC 2151—390873048) and under the project IDs 432325352 – SFB 1454, 450149205 – TRR 333, 528702961 - MESID, and by the University of Bonn (via the Schlegel Professorship of J.H.). 
FTB was supported by LIBIS, and de.NBI, the German Network for Bioinformatics Infrastructure (W-de.NBI-016).
GitHub Copilot was used for code drafting and review only.

\bibliographystyle{abbrvnat}
\bibliography{reference}

@article{SchaelteFro2023,
    author = {Schälte, Yannik and Fröhlich, Fabian and Jost, Paul J and Vanhoefer, Jakob and Pathirana, Dilan and Stapor, Paul and Lakrisenko, Polina and Wang, Dantong and Raimúndez, Elba and Merkt, Simon and Schmiester, Leonard and Städter, Philipp and Grein, Stephan and Dudkin, Erika and Doresic, Domagoj and Weindl, Daniel and Hasenauer, Jan},
    title = {{pyPESTO}: a modular and scalable tool for parameter estimation for dynamic models},
    journal = {Bioinformatics},
    volume = {39},
    number = {11},
    pages = {btad711},
    year = {2023},
    month = {11},
    issn = {1367-4811},
    doi = {10.1093/bioinformatics/btad711},
    eprint = {https://academic.oup.com/bioinformatics/article-pdf/39/11/btad711/53962204/btad711.pdf},
}

@article{SchmiesterSch2021,
	author = {Schmiester, Leonard AND Sch{\"a}lte, Yannik AND Bergmann, Frank T. AND Camba, Tacio AND Dudkin, Erika AND Egert, Janine AND Fr{\"o}hlich, Fabian AND Fuhrmann, Lara AND Hauber, Adrian L. AND Kemmer, Svenja AND Lakrisenko, Polina AND Loos, Carolin AND Merkt, Simon AND M{\"u}ller, Wolfgang AND Pathirana, Dilan AND Raim{\'u}ndez, Elba AND Refisch, Lukas AND Rosenblatt, Marcus AND Stapor, Paul L. AND St{\"a}dter, Philipp AND Wang, Dantong AND Wieland, Franz-Georg AND Banga, Julio R. AND Timmer, Jens AND Villaverde, Alejandro F. AND Sahle, Sven AND Kreutz, Clemens AND Hasenauer, Jan AND Weindl, Daniel},
	date = {26-01-2021},
	doi = {10.1371/journal.pcbi.1008646},
	journal = {PLoS Computational Biology},
	month = {January},
	number = {1},
	pages = {1-10},
	publisher = {Public Library of Science},
	title = {{PEtab}---Interoperable specification of parameter estimation problems in systems biology},
	volume = {17},
	year = {2021},}

@article{HuckaFin2003,
	author = {Hucka, M. and Finney, A. and Sauro, H. M. and Bolouri, H. and Doyle, J. C. and Kitano, H. and Arkin, A. P. and Bornstein, B. J. and Bray, D. and Cornish-Bowden, A. and Cuellar, A. A. and Dronov, S. and Gilles, E. D. and Ginkel, M. and Gor, V. and Goryanin, I. I. and Hedley, W. J. and Hodgman, T. C. and Hofmeyr, J.-H. and Hunter, P. J. and Juty, N. S. and Kasberger, J. L. and Kremling, A. and Kummer, U. and {Le Nov\`{e}re}, N. and Loew, L. M. and Lucio, D. and Mendes, P. and Minch, E. and Mjolsness, E. D. and Nakayama, Y. and Nelson, M. R. and Nielsen, P. F. and Sakurada, T. and Schaff, J. C. and Shapiro, B. E. and Shimizu, T. S. and Spence, H. D. and Stelling, J. and Takahashi, K. and Tomita, M. and Wagner, J. and Wang, J.},
	doi = {10.1093/bioinformatics/btg015},
	journal = {Bioinformatics},
	journal-full = {Bioinformatics},
	months = {March},
	number = {4},
	pages = {524--531},
	title = {The systems biology markup language {(SBML):} {A} medium for representation and exchange of biochemical network models},
	volume = {19},
	year = {2003}}

@article{FroehlichWei2021,
	author = {Fr{\"o}hlich, Fabian and Weindl, Daniel and Sch{\"a}lte, Yannik and Pathirana, Dilan and Paszkowski, {\L}ukasz and Lines, Glenn Terje and Stapor, Paul and Hasenauer, Jan},
	date = {02-04-2021},
	doi = {10.1093/bioinformatics/btab227},
	eprint = {https://academic.oup.com/bioinformatics/advance-article-pdf/doi/10.1093/bioinformatics/btab227/37371345/btab227.pdf},
	issn = {1367-4803},
	journal = {Bioinformatics},
	month = {April},
	title = {{AMICI: high-performance sensitivity analysis for large ordinary differential equation models}},
	volume = {btab227},
	year = {2021},}

@article{RaueSte2015,
	author = {Raue, A. and Steiert, B. and Schelker, M. and Kreutz, C. and Maiwald, T. and Hass, H. and Vanlier, J. and T{\"o}nsing, C. and Adlung, L. and Engesser, R. and Mader, W. and Heinemann, T. and Hasenauer, J. and Schilling, M. and H{\"o}fer, T. and Klipp, E. and Theis, F. J. and Klingm{\"u}ller, U. and Sch{\"o}berl, B. and J.Timmer},
	doi = {10.1093/bioinformatics/btv405},
	journal = {Bioinformatics},
	journal-full = {Bioinformatics},
	months = {Nov.},
	number = {21},
	pages = {3558--3560},
	title = {{Data2Dynamics:} a modeling environment tailored to parameter estimation in dynamical systems},
	volume = {31},
	year = {2015},}

@article {PerssonFro2025,
	author = {Persson, Sebastian and Fröhlich, Fabian and Grein, Stephan and Loman, Torkel and Ognissanti, Damiano and Hasselgren, Viktor and Hasenauer, Jan and Cvijovic, Marija},
    title = {PEtab.jl: advancing the efficiency and utility of dynamic modelling},
    journal = {Bioinformatics},
    volume = {41},
    number = {9},
    pages = {btaf497},
    year = {2025},
    month = {09},
    issn = {1367-4811},
    doi = {10.1093/bioinformatics/btaf497},
}

@article{HoopsSah2006,
	author = {Hoops, S. and Sahle, S. and Gauges, R. and Lee, C. and Pahle, J. and Simus, N. and Singhal, M. and Xu, L. and Mendes, P. and Kummer, U.},
	doi = {10.1093/bioinformatics/btl485},
	journal = {Bioinformatics},
	journal-full = {Bioinformatics},
	number = {24},
	pages = {3067--3074},
	title = {{COPASI} -- a {CO}mplex {PA}thway {SI}mulator},
	volume = {22},
	year = {2006},}

@article{ArmisteadHoe2024,
  title={A sphingolipid rheostat controls apoptosis versus apical cell extrusion as alternative tumour-suppressive mechanisms},
  author={Armistead, Joy and H{\"o}pfl, Sebastian and Goldhausen, Pierre and M{\"u}ller-Hartmann, Andrea and Fahle, Evelin and Hatzold, Julia and Franzen, Rainer and Brodesser, Susanne and Radde, Nicole E and Hammerschmidt, Matthias},
  journal={Cell Death \& Disease},
  volume={15},
  number={10},
  pages={746},
  year={2024},
  publisher={Nature Publishing Group UK London},
  doi={10.1038/s41419-024-07134-2}
}

@article{BurbanoKem2024,
  title={Basal MET phosphorylation is an indicator of hepatocyte dysregulation in liver disease},
  author={Burbano de Lara, Sebastian and Kemmer, Svenja and Biermayer, Ina and Feiler, Svenja and Vlasov, Artyom and D’Alessandro, Lorenza A and Helm, Barbara and M{\"o}lders, Christina and Dieter, Yannik and Ghallab, Ahmed and others},
  journal={Molecular systems biology},
  volume={20},
  number={3},
  pages={187--216},
  year={2024},
  publisher={Nature Publishing Group UK London},
  doi={10.1038/s44320-023-00007-4}
}

@article{MalikGlo2020,
        author = {Malik-Sheriff, Rahuman S and Glont, Mihai and Nguyen, Tung V N and Tiwari, Krishna and Roberts, Matthew G and Xavier, Ashley and Vu, Manh T and Men, Jinghao and Maire, Matthieu and Kananathan, Sarubini and Fairbanks, Emma L and Meyer, Johannes P and Arankalle, Chinmay and Varusai, Thawfeek M and Knight-Schrijver, Vincent and Li, Lu and Dueñas-Roca, Corina and Dass, Gaurhari and Keating, Sarah M and Park, Young M and Buso, Nicola and Rodriguez, Nicolas and Hucka, Michael and Hermjakob, Henning},
        title = "{BioModels — 15 years of sharing computational models in life science}",
        journal = {Nucleic Acids Research},
        volume = {48},
        number = {D1},
        pages = {D407-D415},
        year = {2020},
        month = {1},
        issn = {0305-1048},
        doi = {10.1093/nar/gkz1055},
        eprint = {https://academic.oup.com/nar/article-pdf/48/D1/D407/31698010/gkz1055.pdff},
}

@article{HassLoo2019,
	author = {Hass, Helge and Loos, Carolin and Raim{\'u}ndez-{\'A}lvarez, Elba and Timmer, Jens and Hasenauer, Jan and Kreutz, Clemens},
	doi = {10.1093/bioinformatics/btz020},
	eprint = {https://academic.oup.com/bioinformatics/article-pdf/35/17/3073/29591749/btz020.pdf},
	issn = {1367-4803},
	journal = {Bioinformatics},
	month = {01},
	number = {17},
	pages = {3073-3082},
	title = {{Benchmark problems for dynamic modeling of intracellular processes}},
	volume = {35},
	year = {2019},}

@article{TiwariKri2021,
    author = {Tiwari, Krishna and Kananathan, Sarubini and Roberts, Matthew G and Meyer, Johannes P and Sharif Shohan, Mohammad Umer and Xavier, Ashley and Maire, Matthieu and Zyoud, Ahmad and Men, Jinghao and Ng, Szeyi and Nguyen, Tung V N and Glont, Mihai and Hermjakob, Henning and Malik-Sheriff, Rahuman S},
    title = {Reproducibility in systems biology modelling},
    journal = {Molecular Systems Biology},
    volume = {17},
    number = {2},
    pages = {e9982},
    doi = {10.15252/msb.20209982},
    eprint = {https://www.embopress.org/doi/pdf/10.15252/msb.20209982},
    year = {2021}
}

@article{SmithBer2009,
	author = {Smith, Lucian P. and Bergmann, Frank T. and Chandran, Deepak and Sauro, Herbert M.},
	doi = {10.1093/bioinformatics/btp401},
	eprint = {https://academic.oup.com/bioinformatics/article-pdf/36/2/594/31962762/btz581.pdf},
	journal = {Bioinformatics},
	month = {09},
	number = {18},
	pages = {2452--2454},
	title = {Antimony: a modular model definition language},
	volume = {25},
	year = {2009},}

@article{Bergmann2023,
 doi = {10.21105/joss.05553},
 year = {2023},
 publisher = {The Open Journal},
 volume = {8},
 number = {90},
 pages = {5553},
 author = {Frank T. Bergmann},
 title = {BASICO: A simplified Python interface to COPASI},
 journal = {Journal of Open Source Software}
}

@book{Fitzpatrick2021,
  title={Create GUI Applications with Python \& Qt6 (PySide6 Edition)},
  author={Fitzpatrick, Martin},
  year={2021},
  publisher={Martin Fitzpatrick}
}

@article{VillaverdePat2021,
	author = {Villaverde, Alejandro F and Pathirana, Dilan and Fr{\"o}hlich, Fabian and Hasenauer, Jan and Banga, Julio R},
	doi = {10.1093/bib/bbab387},
	eprint = {https://academic.oup.com/bib/advance-article-pdf/doi/10.1093/bib/bbab387/40534209/bbab387.pdf},
	issn = {1477-4054},
	journal = {Briefings in Bioinformatics},
	month = {10},
	note = {bbab387},
	title = {{A protocol for dynamic model calibration}},
	year = {2021},
}

@article{BergmannAda2014,
	author = {Bergmann, Frank T. and Adams, Richard and Moodie, Stuart and Cooper, Jonathan and Glont, Mihai and Golebiewski, Martin and Hucka, Michael and Laibe, Camille and Miller, Andrew K. and Nickerson, David P. and Olivier, Brett G. and Rodriguez, Nicolas and Sauro, Herbert M. and Scharm, Martin and Soiland-Reyes, Stian and Waltemath, Dagmar and Yvon, Florent and Le Nov{\`e}re, Nicolas},
	citation-subset = {IM},
	completed = {2015-09-01},
	country = {England},
	doi = {10.1186/s12859-014-0369-z},
	issn = {1471-2105},
	issn-linking = {1471-2105},
	journal = {BMC bioinformatics},
	month = dec,
	nlm-id = {100965194},
	pages = {369},
	pii = {s12859-014-0369-z},
	pmc = {PMC4272562},
	pmid = {25494900},
	pubmodel = {Electronic},
	pubstate = {epublish},
	revised = {2019-01-08},
	title = {COMBINE archive and OMEX format: one file to share all information to reproduce a modeling project.},
	volume = {15},
	year = {2014}
}

\end{document}